\documentclass[cits]{PoS}

\usepackage{amsmath}
\usepackage{amssymb}
\usepackage{bbm}

\def \Nt {{N_{\tau}}}
\def \Nc {{N_{c}}}
\def \Nf {{N_{f}}}

\newcommand{\expval}[1]{\left\langle #1 \right\rangle}

\newcommand{\pbp}{\bar{\psi}\psi}

\newcommand{\lr}[1]{\left( #1 \right)}

\newcommand{\nn} {\nonumber\\}

\newcommand{\beqn} {\begin{equation}}
\newcommand{\eqn} {\end{equation}}

\def \beq{\begin{equation}}
\def \eeq{\end{equation}}
\def \bea{\begin{eqnarray}}
\def \eea{\end{eqnarray}}

\def \tr {{\rm tr}}

\def \bet0{\beta_0}
\def \bet1{\beta_1}
\def \simgt{\,\rlap{\lower 7.5 pt\hbox{$\mathchar \sim$}}\raise 3 pt \hbox{$>$}\,}
\def \simlt{\,\rlap{\lower 7.5 pt\hbox{$\mathchar \sim$}}\raise 3 pt \hbox{$<$}\,}

\def\lsim{\raise0.3ex\hbox{$<$\kern-0.75em\raise-1.1ex\hbox{$\sim$}}}
\def\gsim{\raise0.3ex\hbox{$>$\kern-0.75em\raise-1.1ex\hbox{$\sim$}}}

\newcommand{\hnu}{{\hat{\nu}}}

\newcommand{\Ord}[1]{\mathcal{O}\left( #1 \right)}

\title{The Phase Diagram of Strong Coupling QCD including Gauge Corrections}

\ShortTitle{Strong Coupling QCD including Gauge Corrections}

\author{Philippe de Forcrand\\
Institute for Theoretical Physics, ETH Z\"urich, CH-8093 Z\"urich, Switzerland and\\
CERN, Physics Department, TH Unit, CH-1211 Geneva 23, Switzerland\\
        E-mail: \email{forcrand@phys.ethz.ch}}

\author{Jens Langelage\\
Institute for Theoretical Physics, ETH Z\"urich, CH-8093 Z\"urich, Switzerland\\
        E-mail: \email{ljens@phys.ethz.ch}}

\author{Owe Philipsen, \speaker{Wolfgang Unger}\\
Institut f\"ur Theoretische Physik, Goethe-Universit\"at Frankfurt,\\
60438 Frankfurt am Main, Germany\\
        E-mail: \email{philipsen@th.physik.uni-frankfurt.de}\\
	\hspace{11mm} \email{unger@th.physik.uni-frankfurt.de}}

\abstract{
\vspace*{-12.5cm}
\begin{flushright}
\texttt{\footnotesize CERN-PH-TH/2013-272}\\
\end{flushright}
\vspace*{11.5cm}
The strong coupling limit of lattice QCD with staggered fermions has been studied for decades, both via Monte Carlo and via mean field theory.
In this model, the finite density sign problem can be made mild and the full phase diagram can be obtained, even in the chiral limit. 
It is however desirable to understand the effect of a finite lattice gauge coupling $\beta$ on the phase diagram in the $\mu-T$ plane in order to understand how it evolves into
the phase diagram of continuum QCD. Here we discuss how to construct a partition function for non-zero lattice coupling, exact to $\Ord{\beta}$,
and present corresponding Monte Carlo results, in particular for corrections to the chiral susceptibility and to the phase diagram.}

\FullConference{31st International Symposium on Lattice Field Theory - LATTICE 2013\\
		July 29 - August 3, 2013\\
		Mainz, Germany}

\newcommand{\p}{\chi}
\newcommand{\pb}{\bar{\chi}}
\newcommand{\dpdpb}{d\chi d \bar{\chi}}

\begin{document}

\section{Introduction}

\begin{figure}[t!]
 \vspace{-7mm}
\includegraphics[width=0.56\textwidth]{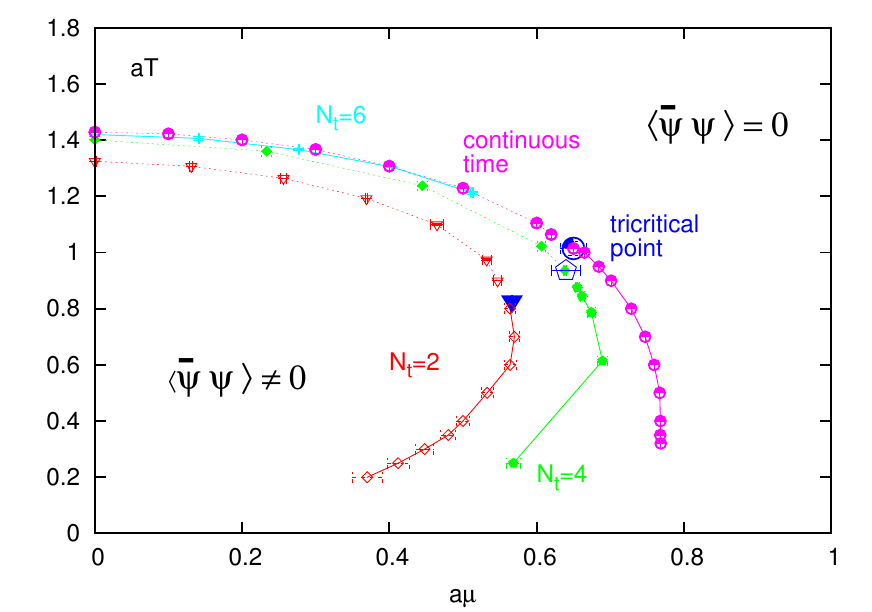}
\includegraphics[width=0.42\textwidth]{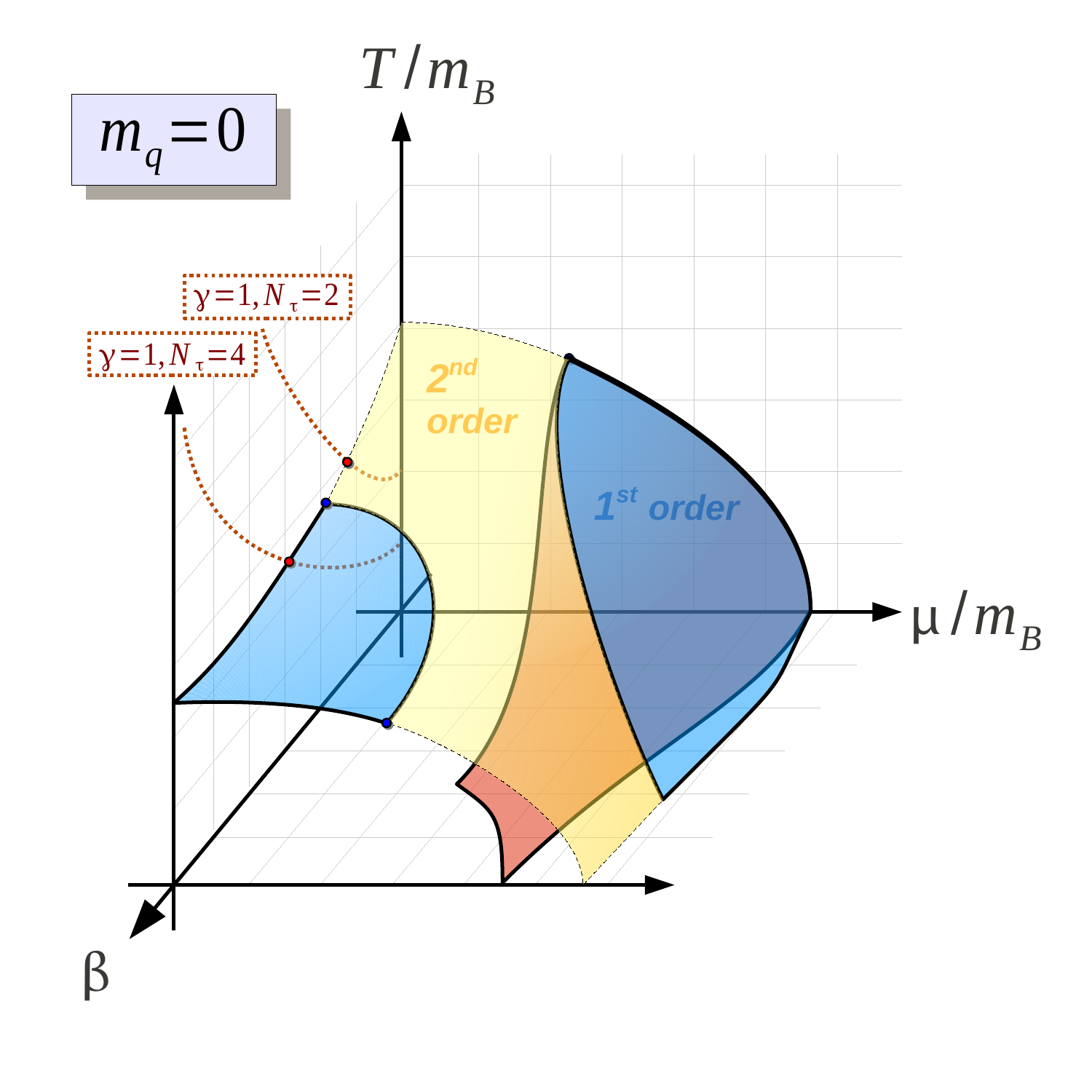}
\vspace{-3mm}
\caption{
\emph{Left:} $\beta=0$ phase diagram with identifications: $aT=\frac{\gamma^2}{\Nt}$, $a\mu=\gamma^2 a_{\tau} \mu$, for $\Nt=2,4,6$ and $\infty$. The $\Nt$-dependence
is mostly caused by deviations from $a/a_\tau=\gamma^2$.
Note that the re-entrance at low temperatures vanishes 
in continuous time ($\Nt\rightarrow\infty$).
\emph{Right:} One of various possible scenarios for the extension of the chiral transition to finite $\beta$. It is expected that the chiral transition and the nuclear transition will split. 
The first and second order transition regions are separated by tricritical lines.
}
\label{PhaseDiagSC}
\label{phasediag}
\end{figure}

It is one of the main goals of lattice QCD at finite temperature and density to map the phase diagram and the order of the transitions as a function of the quark chemical potential 
$\mu$ and the temperature $T$.
However, due to the sign problem of determinant-based Hybrid Monte Carlo (HMC), little progress has been made towards this goal.
All the methods at hand are limited to small $\mu/T$ \cite{Forcrand2009}.
Here we propose to study the phase diagram from a strong coupling perspective, where simulations are feasible also at finite chemical potential.
The strategy in strong coupling lattice QCD is to perform the link integrals analytically first, and then to integrate out the Grassmann variables, hence no fermion determinant arises.
The sign problem poses no problem as it is much milder than with HMC.
We adopt the staggered fermion discretization, where a reformulation in ``dual variables'' can be obtained \cite{Rossi1984}. The full QCD partition function is given by
\begin{eqnarray}
Z_{QCD} &\!\!=\!&\! \!\int \!\!\dpdpb dUe^{S_G+S_F}\!,\qquad
S_G \!=\! \frac{\beta}{2N_c} \! \sum_P \tr[U_P\!+\!U_P^\dagger] \nn
S_F &\!\!=\!& am_q\sum_x\pb_x \p_x + \frac{1}{2}\sum_{x,\nu} \eta_\nu(x) \gamma^{\delta_{\nu 0}}
\left[ \pb_x e^{a_\tau \mu \delta_{\nu 0}} U_\nu(x)\p_{x\!+\!\hat{\nu}} - \pb_{x\!+\!\hat{\nu}} e^{-a_\tau \mu \delta_{\nu 0}}U_\nu^\dagger(x)\p_x \right]
\label{SCQCDPF}
\end{eqnarray}
with $m_q$ the quark mass and $\mu=\frac{1}{3}\mu_B$ the quark chemical potential. The anisotropy $\gamma$ in the Dirac couplings is introduced to vary the temperature continuously.
At strong coupling, the ratio of spatial and temporal lattice spacings is $\frac{a}{a_\tau}\simeq \gamma^2$ \cite{Unger2011}.
The action in the strong coupling limit is simply given by the fermionic action $S_F$, as the lattice gauge coupling $\beta=2\Nc/g^2$ 
vanishes when $g\rightarrow \infty$.
Since the link integration factorizes in the absence of the gauge action, 
the gauge links $U_\nu(x)$ can be integrated out analytically \cite{Eriksson1981}.
After performing the Grassmann integration, the final partition function, introduced in \cite{Rossi1984}, 
is obtained by an analytic rewriting in terms of hadronic degrees of freedom (mesons and baryons):
\begin{align}
Z&=\sum_{\{k,n,\ell\}}\prod_x w_x \prod_{b} w_b \prod_\ell w_\ell \qquad \text{with constraint}\qquad
n_x+\sum_{\hat{\nu}=\pm\hat{0},\ldots \pm \hat{d}}\lr{k_{\hat{\nu}}(x) + \frac{\Nc}{2} |\ell_\hnu(x)|} =\Nc,\nn
w_x&=\frac{\Nc!}{n_x!}(2am_q)^{n_x},\quad w_b=\frac{(\Nc-k_b)!}{\Nc!k_b!}\gamma^{2\delta_{\nu0}},\quad w_\ell=\prod_\ell \lr{\prod_{b\in \ell}\Nc!}^{-1} {\sigma(\ell)} e^{\Nc \Nt r_\ell a_{\tau} \mu}
\gamma^{\Nc N_0}
\end{align}
for gauge group SU($\Nc$). 
The mesons are represented by monomers $n_x\in\{0,\ldots \Nc\}$ on sites $x$ and dimers $k_b\in\{0,\ldots \Nc\}$ on bonds $b=(x,\nu)$, whereas
the baryons are represented by oriented self-avoiding loops $\ell$.
The weight 
$w(\ell)$ of a baryonic loop $\ell$
and in particular its sign $\sigma(\ell)=\pm 1$ depend on the loop geometry.
The admissibility constraint on configurations $\{k,n,\ell\}$ derives from Grassmann integration.
One consequence is that mesonic degrees of freedom (monomers and dimers) can not occupy baryonic sites.
This system has been studied both via mean field \cite{Bilic1992a} and Monte Carlo methods \cite{Karsch1989,Forcrand2010,Unger2011}. In recent years, the Monte Carlo approach has undergone a revival due to the applicability of the worm algorithm.
The idea is to violate the Grassmann constraint locally in order to sample the monomer two-point function $G(x,y)$.
These techniques have been applied to obtain all lattice data presented in this paper. In Fig.~\ref{PhaseDiagSC} (left), we show the ($\mu,T$) phase diagram in the strong coupling limit and for $m_q=0$, 
where $\expval{\pbp}$ is the order parameter for spontaneous chiral symmetry breaking.
It is qualitatively similar to the expected phase diagram of QCD in the chiral limit: the transition is of second order at $a\mu=0$, up to the tricritical point
at $(a\mu_{T},aT_{T})$, where it turns first order. At finite quark mass, the second order line turns into a crossover, the tricritical point into a second order critical endpoint. 
At low temperatures, in contrast to QCD, the chiral transition coincides with the nuclear transition. This is because above the critical chemical potential a baryonic crystal forms,
with one static baryon per spatial site,  
which restores chiral symmetry. This saturation effect is a lattice artifact.

Since strong coupling lattice QCD can be thought of as a one-parameter deformation of \mbox{($\Nf=4$)} continuum QCD, an important question is how both phase diagrams are connected. 
Due to the sign problem, only the plane at $\mu=0$ and the plane at $\beta=0$ are known so far. The QCD phase diagram in the $(\mu,T)$
plane in the continuum limit is largely unknown.
If the $m_q=0$ tricritical point persists in the continuum limit, this is strong evidence for the existence of a chiral critical endpoint in full QCD at physical quark mass.
In order to go beyond the strong coupling limit, we derive a partition function, exact at $\Ord{\beta}$, from which we compute the shift with $\beta$ of the chiral transition temperature.
There are two main questions we want to address: How does the tricritical point move with $\beta$?  And do two distinct transitions (nuclear and chiral) arise at low temperature?
One of various possible scenarios is sketched Fig.~\ref{phasediag} (right).

\section{Corrections to the Strong Coupling Limit}

\begin{figure}[t!]
 \vspace{-8mm}
\centerline{\includegraphics[width=0.9\textwidth]{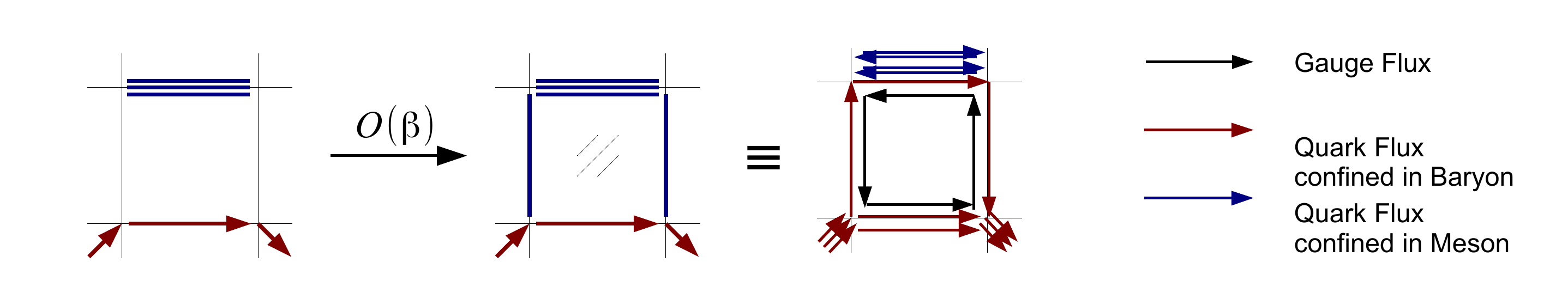}}
 \vspace{-5mm}
\caption{Illustration of $\Ord{\beta}$-corrections to the strong coupling ensemble: insertion of two parallel dimers produces one of the 19 plaquette diagrams. The dimer and flux variables adjacent to the plaquette are 
composed of quark flux and gauge flux: blue lines represent mesonic content, red lines represent baryonic content. The baryon becomes an extended object and interacts with mesons by quark and gluon exchange.
}
\label{diagram}
\end{figure}

To go beyond the strong coupling (SC) limit, a systematic expansion of the QCD partition function in $\beta$ is needed.
Here we derive this expansion valid to the leading order $\Ord{\beta}$. 
The partition function including the gauge part can be written in terms of a fermionic path integral:
\begin{align}
Z_{QCD}=\int d\chi d\bar{\chi} dUe^{S_G+S_F}=\int d\chi d\bar{\chi} Z_{F} \expval{e^{S_G}}_{Z_F}
\end{align}
where $Z_{F}(\chi,\bar{\chi})=\int dU e^{S_F}$ is the fermionic partition function, which is related to the strong coupling ($\beta=0$) partition function via $Z_{SC}=\int d\chi d\bar{\chi} Z_F$.
The weight of the gauge action can then be expressed as an expectation value 
which we linearize to obtain the $\Ord{\beta}$ contribution:
\begin{equation}
\expval{e^{S_G}}_{Z_F}\simeq 1+ \expval{S_G}_{Z_F}=1+\frac{\beta}{2\Nc} \sum_P \expval{\tr[U_P+U_P^\dagger]}_{Z_F}.
 \end{equation}
 To evaluate the expectation value of the elementary plaquette $\expval{\tr[U_P]}_{Z_F}$ in the strong coupling ensemble $Z_F$, we need to compute the link integrals with an additional gauge link coming from
 the plaquette. Before Grassmann integration, the plaquette is given by
 $\expval{\tr[U_P]}_{Z_F}=J_{ij}J_{jk}J_{kl}J_{li}$ with the link integrals at the edges of an elementary plaquette \cite{Creutz1978}:
\begin{align}
 J_{ij}=&\sum_{k=1}^\Nc\frac{(\Nc-k)!}{\Nc!(k-1)!} (M_\chi M_\varphi)^{k-1} \bar{\chi}_j\varphi_i
 -\frac{1}{\Nc!(\Nc-1)!}\epsilon_{i i_1 i_2}\epsilon_{j j_1 j_2} \bar{\varphi}_{i_1}\bar{\varphi}_{i_2} \chi_{j_1}\chi_{j_2}
-\frac{1}{\Nc}\bar{B}_\varphi B_\chi \bar{\chi}_j \varphi_i
 \end{align}
with $M$ and $B$ representing the mesons and baryons.
 From these link integrals, we can compute the weight for inserting a plaquette into the strong coupling configuration. 
At the corners of the plaquette, the Grassmann variables $\varphi,\chi$ 
are bound into baryons and mesons to fulfill a modified constraint: together with plaquette links, the degrees of freedom add up to $\Nc+1$. For $\Nc=3$, there are 19 diagrams contributing to the plaquette $P$, one of them given in Fig.~\ref{diagram}.
We can summarize the modifications of the site weights, $\nu$, and of the
link weights, $\rho$, around an excited plaquette as:
\begin{align}
\nu_M&=(\Nc-1)!,&  \nu_B&=\Nc!,& \rho_{M_k}&=k,&
 \rho_{B_1}&=\frac{1}{(\Nc-1)!},& \rho_{B_2}&=(\Nc-1)!,
\end{align}
for a site connected to a mesonic or a baryonic "external leg",
and for an unoriented [mesonic] or oriented [baryonic] link, respectively.
$B_1$ and $B_2$ represent color singlets $qqg$ and $B\bar{q}g$.
We can then introduce a new set of variables, the plaquette occupation
numbers $q_P \in \{0,1\}$\footnote{
An excited plaquette $q_P=1$ enforces site numbers $q_x=1$ at its 4 corners, and along its 4 edges bond numbers $q_b=1$ for unoriented
bonds and $q_B=1$ for oriented link states $B_1, B_2$.
Otherwise, $q_x=q_b=q_B=0$.
}
and include a Metropolis update allowing to sample the partition function 
\begin{align}
Z(\beta)&=\sum_{\{k,n,q,\ell\}}\prod_x \hat{w}_x \prod_{b} \hat{w}_b \prod_\ell \hat{w}_\ell \prod_P {\hat{w}_P},\nn
\hat{w}_x&=w_x {\nu_i}^{q_x},\qquad \hat{w}_b=w_b {\rho_{M_k}}^{q_b}\qquad \hat{w}_\ell=w_\ell \prod_{B_j \in \ell} { \rho_{B_j}}^{q_B},\qquad
{\hat{w}_P=\left(\frac{\beta}{2\Nc}\right)^{q_P}}
\end{align}
at finite $\beta$.
Qualitatively new aspects of the $\Ord{\beta}$ contributions are 
(\emph{i}) that mesons and baryons are now allowed to interact by quark and gluon exchange and 
(\emph{ii}) that hadrons become extended objects, in contrast to their pointlike nature in the strong coupling limit.
Due to the plaquettes, there is no strict decomposition of the lattice into mesonic and baryonic sites. 
The $\Ord{\beta}$ corrections allow to measure the $\beta=0$ vev of gauge observables (average plaquette, Polyakov loop), 
and the $\Ord{\beta}$ vev of fermionic observables (derivative of the chiral susceptibility,
baryon density).

\section{Gauge Observables}

\begin{figure}[t!]
\vspace{-7mm}
\includegraphics[width=\textwidth]{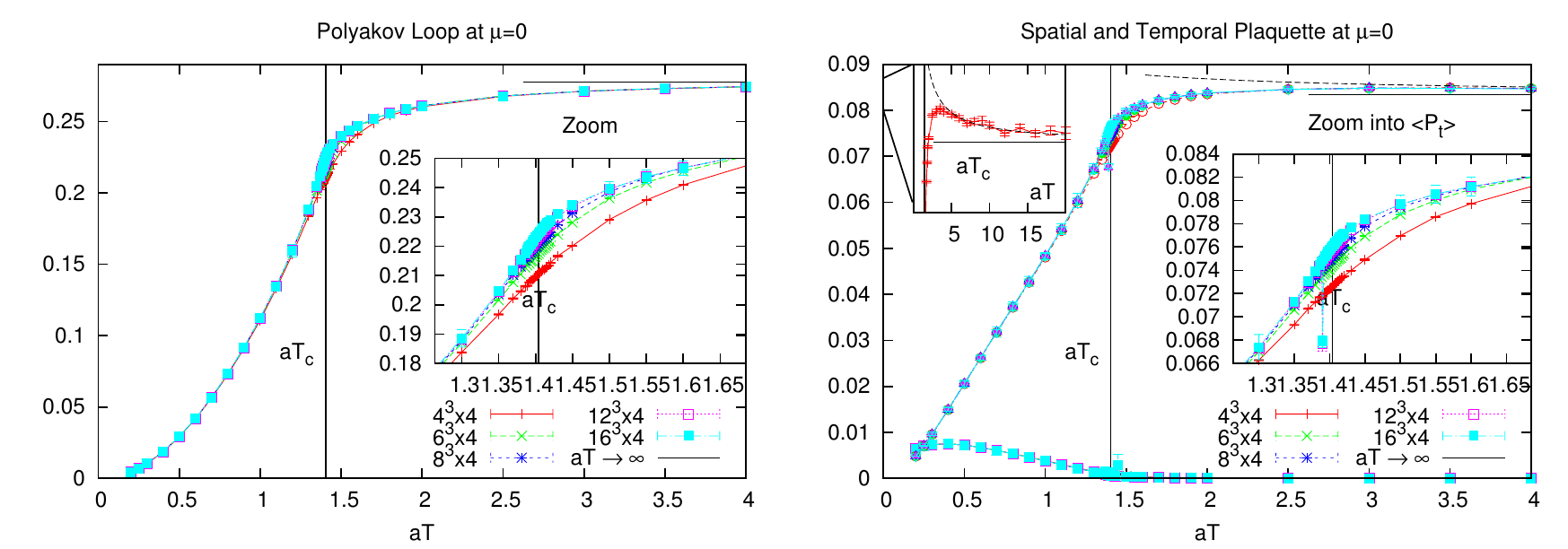}
\vspace{-5mm}
\caption{
Temperature dependence of gauge observables: both the Polyakov loop \emph{(left)} and the plaquette \emph{(right)} show an $L$-dependence in the transition region, close to $aT_c=1.402(1)$.
Note that the temporal plaquette shows a non-monotonic convergence to the high-temperature limit (shown in the upper left inset).
}
\label{polgauge}
\end{figure}

As a byproduct, the above method yields the plaquette expectation value at $\beta=0$. And it can be straightforwardly modified, replacing the plaquette by an arbitrary Wilson loop. 
In this manner, we have obtained the Polyakov loop and plaquette (spatial and temporal) expectation values at $\beta=0$. Our results are shown in Fig. \ref{polgauge} for $\mu=0$. They agree with $\mu=0$ HMC. Note the
appearance of a cusp in both observables as the volumes increases at the second order chiral transition.
This behaviour was reported for U(3) gauge theory in \cite{Fromm2011}. It is not directly related to deconfinement, but to the singularity of the free energy.\footnote{
Including the chemical potential (see Fig. \ref{Miura}, left), the limits for large $\rho=a(T^2 + \mu^2)^{1/2}$ for the (anti-)Polyakov loop $\expval{L^{\pm}}$ and the temporal plaquette
$\expval{P_t}$ are
$$\lim_{\rho\rightarrow \infty} \expval{L^{\pm}}=\frac{1}{\Nc}\frac{\exp(\pm \Nc \mu/T)+\Nc}{2\cosh(\Nc\mu/T)+\Nc+1},\qquad  
\lim_{\rho\rightarrow \infty} \expval{P_t}=\frac{1}{\Nc^3}\left(\frac{\Nc(1+{\frac{1}{2\gamma^2}})\exp(\Nc\mu/T)+\frac{\Nc(\Nc+1)}{2}}{2\cosh(\Nc\mu/T)+\Nc+1}\right)^2.$$
The $\gamma$-dependence of the plaquette is a finite $\Nt$ effect and produces a non-monotonic behaviour, shown in Fig. \ref{polgauge}, right.
}

\section{Phase Diagram as a Function of $\beta$}
For fermionic observables, such as the chiral susceptibility or the baryon density, we can extract the leading order $\beta$-correction (the derivative with respect to $\beta$).
This allows us to determine the $\Ord{\beta}$ gauge corrections to the strong coupling phase diagram.
In the following we consider the chiral limit, where the chiral condensate is zero due to the finite system size, but the susceptibility still signals the chiral transition.
The worm algorithm samples the 2-point correlation function in the 2-monomer sector; its integral is the susceptibility
$\chi_=\frac{1}{V}\sum_{x_1,x_2} G(x_1,x_2) \equiv \expval{(\pbp)^2}$.
The leading order Taylor coefficient of $\chi$, i.e.~$c_{\chi}=\left.\frac{\partial \chi}{\partial \beta}\right|_{\beta=0}$, obeys (with $P=\expval{\frac{1}{2\Nc}\tr[U_P+U_P^\dagger]}_{Z_F}$) \footnote{
At finite temperature, since the spatial and temporal plaquette differ, there are two Taylor coefficients, 
$c_{s}$, $c_{t}$, with $c_{s,t}=\frac{\partial}{\partial \beta_{s,t}}\expval{(\pbp)^2}$. However, $c_{s}$ is largely suppressed with temperature, 
just like the spatial plaquette itself (see Fig.~\ref{polgauge} right), and can be safely neglected.
}
\begin{align}
\chi(\beta) &= \chi_0+c_{\chi}\beta + \Ord{\beta^2},\quad c_{\chi}=3N_s^3 N_t\lr{ \expval{(\pbp)^2 P }-\expval{(\pbp)^2}\expval{P}}.
\end{align}
\begin{figure}[t!]
\vspace{-7mm}
\centerline{\includegraphics[width=0.95\textwidth]{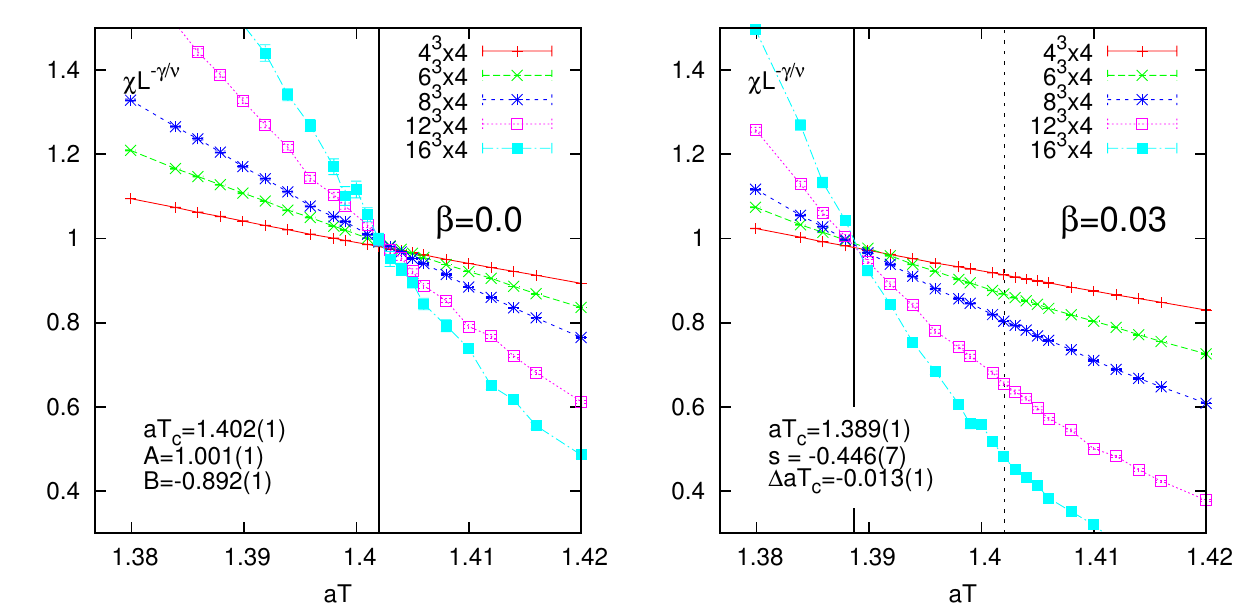}}
\vspace{-3mm}
\caption{
The transition temperature from critical scaling of the chiral susceptibility. \emph{Left:} for $\beta=0$, $aT_c=1.402(1)$. \emph{Right:}
for $\beta=0.03$, the transition temperature shifts to $aT_c=1.389(1)$.}
\label{susc}
\vspace{-8mm}
\end{figure}

\begin{figure}[t]
\hspace{-7mm}
\includegraphics[width=0.59\textwidth]{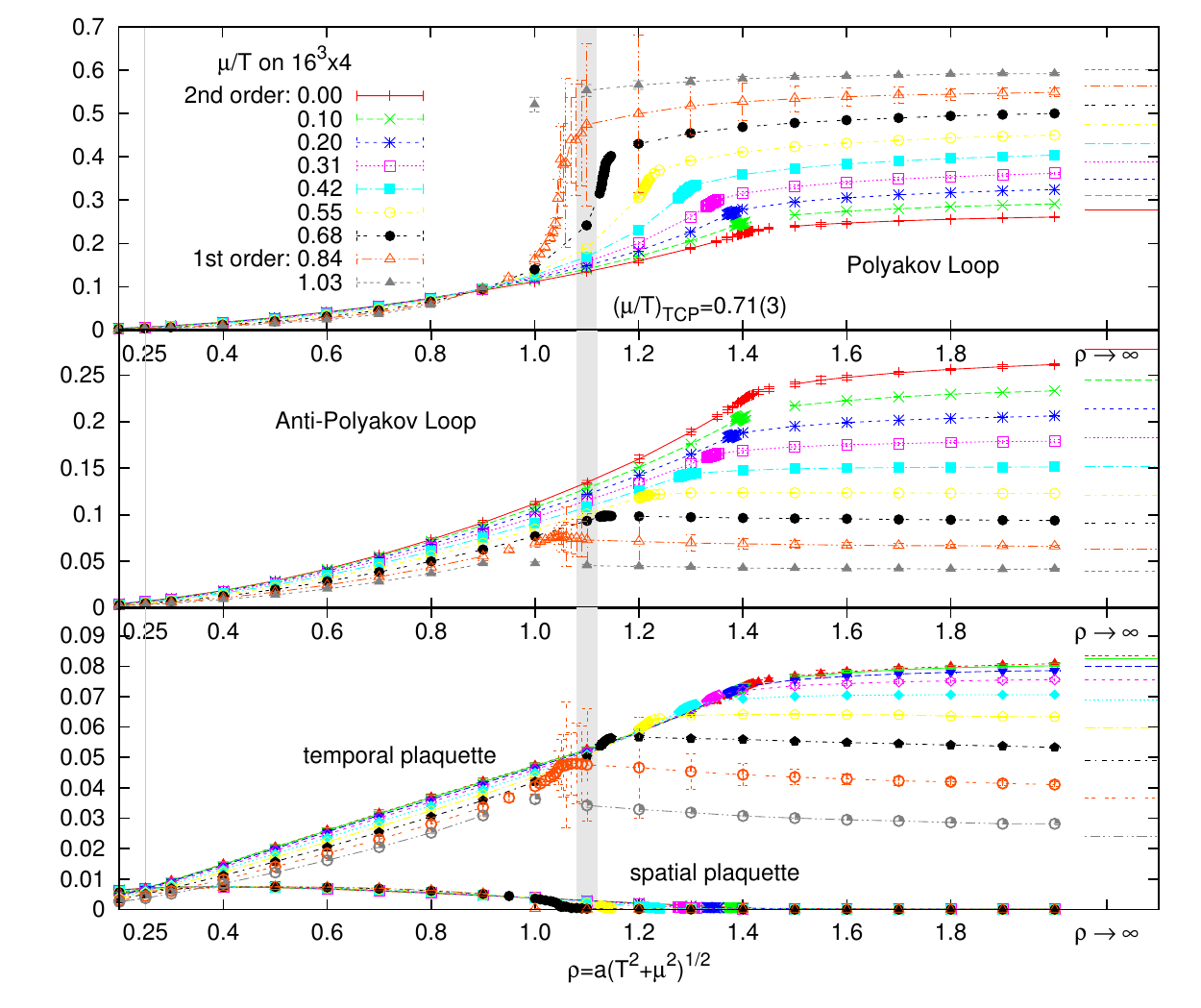}
\includegraphics[width=0.47\textwidth]{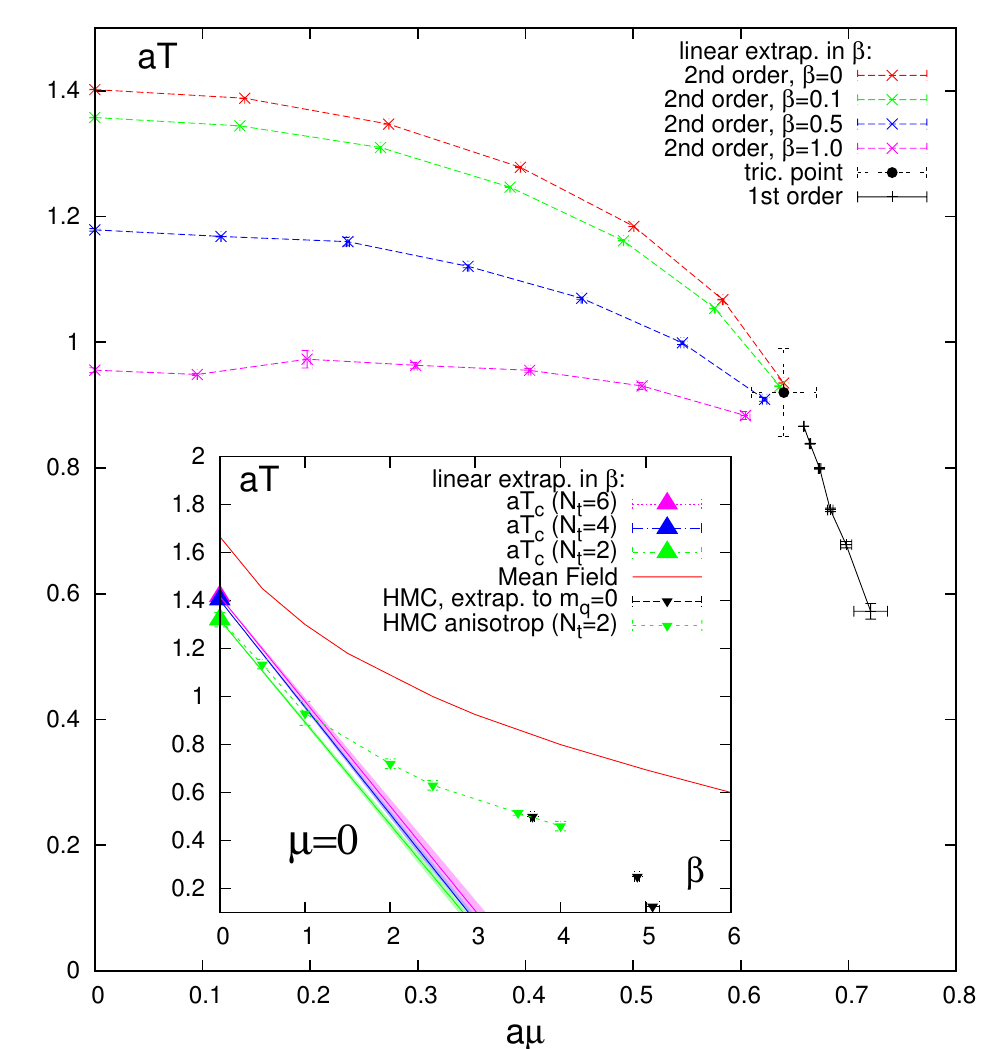}
\caption{
\emph{Left:} The (anti-)Polyakov loop and spatial/temporal plaquette as a function of $a\sqrt{\mu^2+T^2}$, for different values of $\mu/T$. 
\emph{Right:} corrections to the phase boundary when linearly extrapolating in $\beta$ (for $\Nt=4$). \emph{Inset:} 
Comparison of $aT_c(\mu=0)$ versus $\beta$ for $\Nt=2,4,6$ with those obtained by mean field theory \cite{Miura2009} and by HMC (green: anisotropic lattices,
black: isotropic lattices). The $\Nt$ dependence of the extrapolation is small, and is well under control for $\beta<1$.
}
\label{Miura}
\end{figure}

\noindent We determine the transition temperature via critical scaling with 3d O(2) critical exponents $\gamma$, 
$\nu$:
$\chi_L(T,\beta)/L^{\gamma/\nu}=A+B t L^{1/\nu}$, $t=\frac{T-T_c(\beta=0)}{T_c(\beta=0)}$. 
The chiral susceptibility data collapses on a universal scaling function when rescaled in this way, which is almost linear in the scaling window
with non-universal coefficients: $A\simeq 1.001(5)$ and $B\simeq -0.892(5)$ for SU(3) at zero density.
Our strategy is to determine the shift in $aT_c$ induced by a small, finite value of $\beta$.
For the data collapse to be preserved, the Taylor coefficient $c_\chi$ also has to obey critical scaling.\footnote{We find that $c_\chi$ 
can be well fitted by a linear function in $t$: $ \frac{c_\chi}{\chi} \simeq c_1+c_2L^{1/\nu}+c_3t$,
with $c_2=-0.248(1)$ at $\mu=0$.}
The slope of the critical temperature is
$s\equiv\left.\frac{d}{d\beta}aT_c(\beta)\right|_{\beta=0}=-aT_c\frac{A}{B}c_2.$
At $\mu=0$, where $aT_c =1.402(1)$, we obtain $s=-0.446(7)$.
We obtain a consistent result by reweighting to non-zero $\beta$ (see Fig.~\ref{Miura} right).
We can then compare the resulting $\left.aT_c(\beta)\right|_{\mu=0}$ to the mean field result of Miura et al.~\cite{Miura2009}, see inset of Fig. \ref{Miura} right.
The drop in $aT_c$ is expected since the lattice spacing $a(\beta)$ shrinks as $\beta$ is increased.
Due to the mild sign problem of the dimer representation, our method to determine the slope of $aT_c$ can be readily extended to finite density, with Fig. \ref{Miura} (right) showing
the phase boundary for small $\beta$. We find that the change with $\beta$ weakens with $\mu$ and eventually vanishes at the tricritical point. We do not find any shift in the
first order phase boundary, as sketched in Fig. \ref{phasediag} right. This may change if we include $\Ord{\beta^2}$ corrections, which we plan to do in the future.

\section{Conclusion} 
We have presented a method to compute gauge corrections to the QCD phase diagram at strong coupling.
We make use of reweighting to obtain the leading order gauge corrections to the chiral susceptibility. 
Via a finite size scaling analysis we are able to determine the derivative of the chiral transition temperature $\frac{d}{d\beta}aT_c$, at zero and non-zero density,
resulting in the finite-$\beta$ phase diagram in Fig.~\ref{Miura} right.
In the strong coupling limit, the ratio $\frac{T_c(\mu=0)}{3\mu_c(T=0)}\approx \frac{1.403}{1.71}=0.82$
is much too large compared to the continuum result (in the chiral limit)
$\frac{T_c}{3\mu_c}\approx \frac{154\, \rm MeV}{0.93\, \rm GeV}=0.165$. Hence it is expected that the critical temperature $aT_c(\mu=0)$ decreases more
rapidly with $\beta$ than $a\mu_c(T=0)$.

\emph{Acknowledgements} - We would like to thank K.~Miura and A.~Ohnishi for helpful discussions.
This work is supported by Helmholtz International Center for FAIR. WU and JL thank the SNF for support under grant 200020\_137920.

\end{document}